# Artificial helical nanomagnets

A.A.Fraerman[1], B.A.Gribkov[1], S.A.Gusev[1], B. Hjorvarsson[2], A.Yu.Klimov[1], V.L.Mironov[1], D.S.Nikitushkin[1], V.V.Rogov[1], S.N.Vdovichev[1], H.Zabel[3]

[1]Institute for Physics of Microstructures RAS, Nizhny Novgorod, Russia
[2]Uppsala University, Uppsala, Sweden
[3]Ruhr-Universitet, Bochum, Germany
Electronic mail: andr@ipm.sci-nnov.ru

We demonstrate the existence of a helical state in patterned multilayer nanomagnets. The artificial helimagnets consist of a stack of single domain ferromagnetic disks separated by insulating nonmagnetic spacers. The spiral state is shown to originate from the magnetostatic interaction between nearest and next nearest neighbouring single-domain disks in the stack. Since the magnetostatic interaction for such disks is rather strong, the helical state should be stable at room temperature. The helimagnets were fabricated from a [Co/Si]×3 multilayer using electron beam lithography. The helical state was confirmed by a magnetic stray field analysis using magnetic force microscopy.

PACS numbers: 68.37.Rt, 75.75.+a, 75.70.Cn

Helical magnets are of continued interest because of their unusual thermodynamic, optical, and magneto transport properties [1-9]. Helical magnetic ordering has previously been observed in natural crystals. For instance, a helical magnetization distribution with a spiral vector perpendicular to the close–packed planes is formed in holmium single crystals at temperatures below 133 K [1]. As a rule, helicoidal ordering in crystals appears to result from long-range interactions between the localized magnetic moments of atoms. In this Letter, we will show that a helical magnetization distribution can also be realized in artificial magnetic nanostructures.

The possibility of creating helical nanomagnets is based on the universal long-range interaction between particles through magnetostatic stray fields. The value and sign of the interaction energy for two uniform magnetized particles are determined by mutual orientation of their magnetic moments with respect to the axis connecting these particles. In particular, for two single-domain circular ferromagnetic disks separated by a nonmagnetic spacer, the interaction energy leads to antiferromagnetic ordering (Fig. 1a, b).

Let us now consider the effect from adding a third layer, as illustrated in Fig. 1c, d. The interaction between neighbouring disks favours collinear antiferromagnetic ordering, but leads to frustration in orientation of magnetic moments of the first and third disks. If the interaction between these disks is large enough and if the magnetic moments are confined and free to rotate in the plane, the ground state of this system is noncollinear (Fig. 1d). The magnetostatic energy of the three circle disks can be represented as:

$$E = \varepsilon_{12} Cos\theta_{12} + \varepsilon_{23} Cos\theta_{23} + \varepsilon_{13} Cos\theta_{13}, \quad (1)$$

where $\varepsilon_{ij}$ ($i,j = 1, 2, 3$) are the interaction energy between $i$ and $j$ disks ($\varepsilon_{ij} > 0$), $\theta_{ij}$ is the angle between the magnetic moment directions in the $i$ and $j$ disks ($\theta_{13} = \theta_{12} + \theta_{23}$). In a system of three identical magnetic disks with identical spacers, $\varepsilon_{12} = \varepsilon_{23} = \varepsilon$. Varying expression (1) we obtain a minimum of the magnetostatic energy for $\theta_{23} = \theta_{12} = \theta$, where angle $\theta$ is defined by one of the following equations:

$$Sin\,\theta = 0 \quad (2)$$

$$Cos\,\theta = -\frac{\varepsilon}{2\varepsilon_{13}}. \quad (3)$$

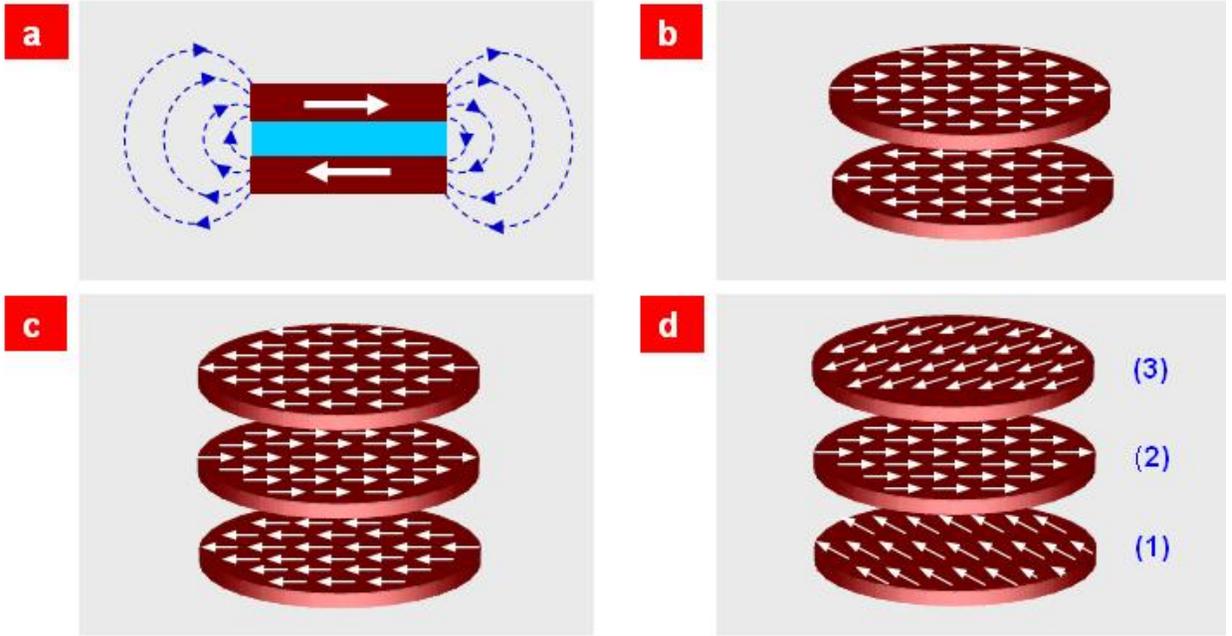

FIG. 1. Magnetic states in multilayer nanomagnets. **a-b**, Schematic pictures of the stray magnetic field (**a**) and antiferromagnetic ordering (**b**) in two single-domain nanodisks. **c-d**, Antiferromagnetic ordering (**c**) and noncollinear magnetic helix (**d**) in the three disks system.

Thus, if the interaction energy of the next nearest neighbours is small, i.e. $2\varepsilon_{13} < \varepsilon$, then an antiferromagnetic ordered (antiparallel) state ($\theta = \pi$) is formed (Fig. 1c). When $2\varepsilon_{13} > \varepsilon$, a non-collinear magnetic spiral state is realized (Fig. 1d). This state is doubly degenerate i.e. the energies for "left" and "right" helicoids are equal.

The main factor determining the presence of noncollinear ordering is the ratio of the interaction between the nearest and the next nearest disks, as seen in equation (3). Taking into account the cylindrical symmetry of the stacked disks and the homogeneity of the magnetization, the energy of the magnetostatic interaction between disks can be represented as:

$$\varepsilon_{ij}(|z_i - z_j|) = M_s V \langle H_{ij} \rangle, \qquad (4)$$

where $M_s$ and $V$ are the saturation magnetization and disk volume, respectively, $z_i$ is the coordinate for the centre of disk $i$, and $\langle H_{ij} \rangle$ is the volume averaged longitudinal (parallel to the magnetization direction) component of the magnetic field induced by disk $i$ within disk $j$.

The results of numerical micromagnetic calculations of the dependence of $\langle H_{ij} \rangle$ on distance $z_i$ for disks with a diameter of 300 nm and a height of 5 nm are presented in Fig. 2.

The numerical calculations show that, for the multilayer nanomagnets with 3 nm thick nonmagnetic spacers, the ratio $\varepsilon/\varepsilon_{13}$ is about 1.25. Thus, using only the magnetostatic interaction between the particles, a helical state with an angular displacement close to *2π/3* is obtained. In contrast to natural helimagnets, the spiral state for three particles is solely caused by next nearest neighbour disk interactions, due to an edge effect. Since the magnetostatic interaction for such disks is rather strong ($\varepsilon_{13} \sim 10^{-10} - 10^{-11}$ erg), the helical state should be stable at room temperature.

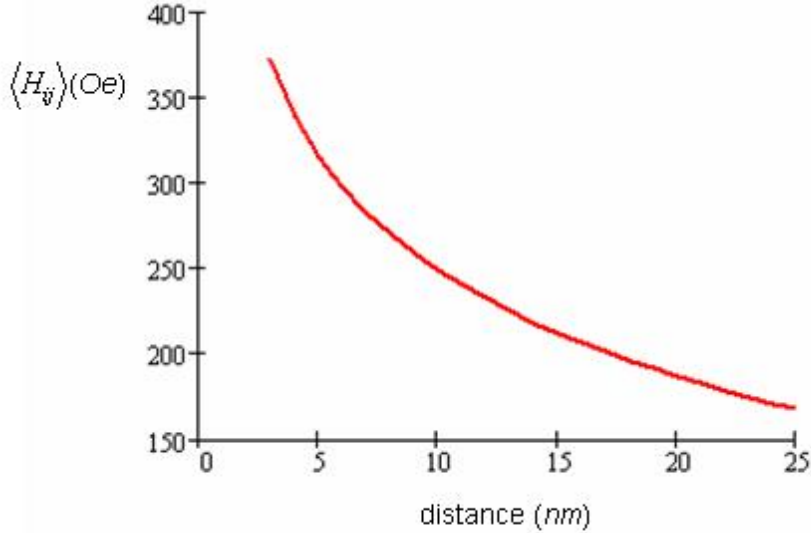

FIG. 2. Dependence of volume averaged longitudinal component of the magnetic field with the distance between the disks.

Our simulations show that when the number of layers (*n*) *is* increased to 4, the ground state is a collinear structure. Moreover, a collinear magnetic structure is the ground state of the system for all even values of *n*. This has been demonstrated for ten magnetic disks with a diameter of 300 and 500 nm [10]. When the number of layers is odd, the ground state is non-collinear. The angle between the magnetic moments of the nearest disks can be approximated by $\sim \pi (n-1)/n$. When the number of layers is large, the spiral state is close to an antiferromagnetic state ($\theta \sim \pi$). Consequently, the system consisting of 3 disks is optimal for creating and observing a helical state.

To verify these concepts, we fabricated multilayer nanomagnets with parameters discussed above, and measured the magnetic stray field distribution of such structures. The Co/Si multilayer nano stacks were fabricated by means of electron beam lithography and ion etching of a multilayer thin film structure grown by magnetron sputtering. The details of the lithography process are described in [11]. The film thicknesses in the structure were controlled by X-ray reflectometry. The coercivity of the ferromagnetic Co films did not exceed 20 Oe. Si layers were used as spacer layers, effectively hindering any interaction except the magnetostatic interaction. The Co disk dimensions (diameter and thickness) were chosen according to the single domain requirement. It is known [12] that single domain state in sufficiently thin ($h \ll h_c$) ferromagnetic disks is the ground state for disk with diameter $d \gg d_c$ (where $h_c$ and $d_c$ are similar to the exchange length). In our experiments, we used disks with diameters of 300 nm and thicknesses less than 20 nm. The magnetic force microscopy (MFM) method was used to observe the helical magnetic state in multilayer nanomagnets (a multimode scanning probe microscope "Solver-HV" (NT-MDT) was used). However, the upper layer mostly contributes to tip-sample interactions and thereby dominates the MFM contrast. Computer modelling shows that the MFM image of the helical state in tri-layer nanomagnets with equal Co thicknesses appears similar to the image of a single uniformly magnetized disk (Fig. 3a). To obtain a clear spiral signature in the MFM contrast we considered a structure where the thickness of the Co layers increases with increasing distance between a layer and the MFM probe. Thus, the change of magnetic moment of the layers was used to compensate for the different distances between the disks and the tip. A structure with Co layers thicknesses of 16, 11, and 8 nm with 3 nm spacers yields optimal contrast for the MFM imaging of the helical state, as judged from these simulations (Fig. 3b). This structure is characterized by a spiral distribution which looks like the well-known eastern symbol Yin-Yang (Fig. 3c).

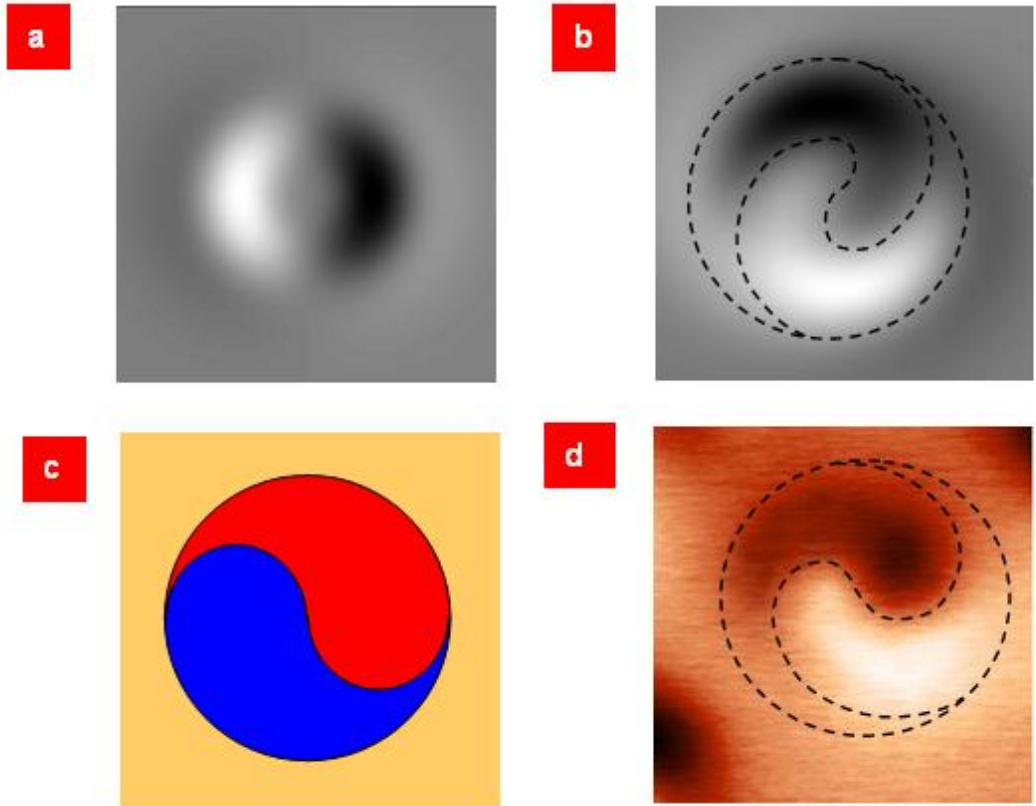

FIG. 3. MFM contrast distributions for helical state. **a**, Simulated MFM contrast for a triple nanodisk with equal Co thicknesses. **b**, Simulated MFM contrast distribution for optimized triple nanodisk. **c**, The Yin-Yang symbol. **d**, Experimental MFM image from triple 16, 11, 8 nm Co nanodisk. (Frame size 1×1 μm). The dashed lines separate the regions with dark and bright contrast to emphasize the spiral symmetry of MFM contrast.

The uniform single-domain state of the disks is essential for the stabilization of the spiral state. The magnetic energy arising from shape anisotropy must be smaller than the magnetostatic interaction. Hence, for the formation of helical state, the disk ellipticity may not exceed $0.8 < a/b < 1.2$, where *a* and *b* are the lateral dimensions of an elliptical disk.

Taking all these conditions into account, tri-layer nanomagnets with Co layer thicknesses of 16, 11, and 8 nm with 3 nm spacers were fabricated and measured. Fig. 3d shows a MFM image for one of the nanomagnets. As seen in this Figure, the MFM contrast image is very similar to the calculated image (Fig. 3b) for the same tri-layer structure in a helical state. As mentioned previously, "left" and "right" helices have equal energies and are therefore equally probable. Figure 4 shows the MFM image of two nanomagnets that demonstrate "left" and "right" handed spiral contrast corresponding to the two different helical states.

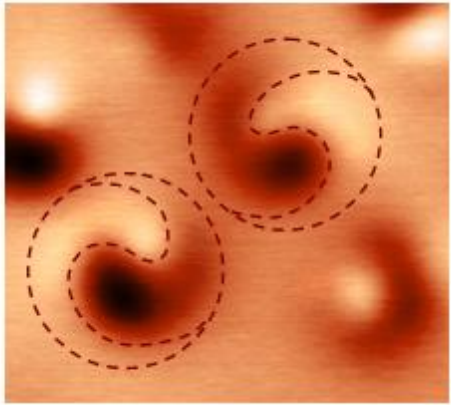

FIG. 4. MFM image of two helical disks with different chirality. The frame size is 1.8×1.8 μm.

In conclusion, artificial helical nanomagnets were fabricated. The helimagnets consist of three single-domain ferromagnetic disks separated by nonmagnetic spacers. The spiral state in this system originates from the magnetostatic interaction between nearest and next nearest neighbouring ferromagnetic disks. It was shown that a spiral state occurs if the energy of magnetostatic interaction between next nearest neighbouring disks is sufficiently large compared to the interaction energy of neighbouring disks. A spiral state emerges due to a slow decrease of the stray magnetic field with distance between the ferromagnetic disks (Fig. 2). Note that in the case of point magnetic dipoles, the ratio $\varepsilon/\varepsilon_{13}$ is equal to 8 and a spiral state is impossible, thus the aspect ratio of the disks is essential for the realisation of a helical state.

Multilayer nanomagnets with optimized parameters were fabricated by electron beam lithography from a multilayer Co/Si thin film structure. Magnetic force microscopy method was applied for observing the helical state. Furthermore, in triple Co/Si nanomagnets, spiral states with different chirality are realized.

The realization of helical nanomagnets is not only of interest for basic investigations of nanomagnets, but they could also be used in spintronic devices. Under an external magnetic field perpendicular to the ferromagnetic layers, the magnetization distribution can be transformed from non-collinear into a non-coplanar cone magnetic spiral. In this case, we expect unusual transport properties along the magnetic field direction due to the breaking of the time reversal symmetry for electrons moving in non-coplanar inhomogeneous magnetic fields [13-16].

The authors are very thankful to I. Nefedov and I. Shereshevsky for assistance and fruitful discussions. This work was supported by the RFBR (05-02-17153, 07-02-01321), INTAS (# 03-51-4778), SFB 491 and by EC through the NANOSPIN project (contract NMP4-CT-2004-013545).